\documentclass[a4paper,11pt]{article}

\usepackage{graphicx,amssymb,amstext,amsmath,mathabx,mathtools,esvect,xparse}
\usepackage{xcolor,wrapfig}
\usepackage{floatflt}
\usepackage{lipsum,ragged2e}
\usepackage{ragged2e,graphicx}
\usepackage[a4paper, total={6.5in, 9in}]{geometry}

\definecolor{purple_nice}{rgb}{0.4,0.2,0.7}
\definecolor{fuel_blue}{RGB}{42,162,185}
\definecolor{YInMn_blue}{RGB}{46, 80, 144}
\definecolor{ultramarine}{RGB}{63, 0, 255}
\definecolor{KLEIN_blue}{rgb}{0, 0.18, 0.65}

\usepackage[linktocpage=true]{hyperref}
\hypersetup{
    colorlinks=true,
    linkcolor=YInMn_blue,
    citecolor=ultramarine,
    filecolor=fuel_blue,
    urlcolor=KLEIN_blue,
}

\usepackage{cite}

\newcommand{\bc}{\begin{center}}
\newcommand{\ec}{\end{center}}
\def\ba#1{\begin{array}{#1}\displaystyle}
\newcommand{\ea}{\end{array}}

\newcommand{\beq}{\begin{equation}}
\newcommand{\eeq}{\end{equation}}
\newcommand{\beqa}{\begin{eqnarray}}
\newcommand{\eeqa}{\end{eqnarray}}

\newcommand{\bi}{\begin{itemize}}
\newcommand{\ei}{\end{itemize}}

\newcommand{\bra}{\langle}
\newcommand{\ket}{\rangle}

\newcommand{\TTb}{\mathrm{T}\overline{\mathrm{T}}}
\newcommand{\bal}{\boldsymbol{\alpha}}
\newcommand{\bol}{\boldsymbol{0}}
\newcommand{\bel}{\boldsymbol{\beta}}

\begin{document}

\begin{titlepage}
\title{Boundary Quantum Field Theories Perturbed by $\TTb$: Towards a Form Factor Program}
\author{Olalla A. Castro-Alvaredo{\color{red} {$^\heartsuit$}}, Stefano Negro{\color{green} {$^\diamondsuit$}} and Fabio Sailis{\color{blue} {$^\clubsuit$}}}
\date{\small {\color{blue} {$^\clubsuit$}}  {\color{red} {$^\heartsuit$}}  Department of Mathematics, City St George's, University of London,\\ 10 Northampton Square EC1V 0HB, UK\\
\medskip
{\color{green} {$^\diamondsuit$}}  Department of Mathematics, University of York,
Heslington, York YO10 5DD, UK\\
\medskip
}
\maketitle
\begin{abstract} 

Our understanding of irrelevant perturbations of integrable quantum field theories has greatly expanded over the last decade. In particular, we know that, from a scattering theory viewpoint at least, their effect is realised as a modification the two-body scattering amplitudes by a CDD factor. While this sounds like a relatively small change, this CDD factor incorporates a non-trivial dependence on the perturbation parameter(s) and alters substantially the high-energy physics of the model. This occurs through the introduction of a natural length scale and is associated with phenomena such as the Hagedorn transition. In this paper we discuss how all these features extend to boundary integrable quantum field theories and propose a construction for the building blocks of matrix elements of local fields. We show that the same type of building blocks are also found in the sinh-Gordon model with Dirichlet boundary conditions.

\end{abstract}


\vfill

\noindent 
{\Large {\color{red} {$^\heartsuit$}}}o.castro-alvaredo@city.ac.uk\\
\noindent 
{\Large {\color{green} {$^\diamondsuit$}}}stefano.negro@york.ac.uk\\
{\Large {\color{blue} {$^\clubsuit$}}}fabio.sailis@city.ac.uk
\hfill \today
\end{titlepage}

\maketitle

\section{Introduction}
 Given an integrable quantum field theory, it has been known since 2016 that a perturbation by $\TTb$ and higher spin versions thereof introduces a deformation of the two-body scattering matrix \cite{VEVTTb,Smirnov:2016lqw}. If the theory is diagonal, that is there is no back-scattering, then the deformation takes the form
 \beq 
S^{\bal}_{ab}(\theta):=S_{ab}(\theta) \Phi^{\bal}_{ab}(\theta)\quad \mathrm{with} \quad \log \Phi^{\bal}_{ab}(\theta)=-i \sum_{s\in \mathcal{S}} m_a^s m_b^s \alpha_s \sinh(s\theta)\,,
\label{s}
 \eeq
 where $S_{ab}(\theta)$ is the original $S$-matrix associated to the process $a+b \mapsto a+b$, with $a, b$ particle quantum numbers. The masses of these particles are denoted by $m_a, m_b$ and $\alpha_s$ are couplings such that the combinations $m_a^s m_b^s \alpha_s$ are dimensionless. The values of $s$ are drawn from the set $\mathcal{S}$ of (integer) spins of local conserved quantities in the IQFT. The bold symbol $\bal$ indicates the set of parameters $\alpha_s$ in the sum. 

 Starting from this $S$-matrix, various techniques commonly associated with integrable quantum field theories (IQFTs) have been applied to $\TTb$-perturbed models. This includes the thermodynamic Bethe ansatz (TBA) \cite{Dubovsky:2012wk, Caselle:2013dra, Cavaglia:2016oda,Dubovsky:2018bmo,Camilo:2021gro,Cordova:2021fnr}, and, most recently, the form factor program \cite{PRL,longpaper,ourentropy,theirentropy,MFF}. 
 If we put our focus on the scattering theory of these models, and on the traditional pathway to studying IQFTs, it is natural to also consider the effect of a $\TTb$ perturbation and its generalisations on IQFTs in the presence of integrable boundaries. The study of this problem was initiated in \cite{Jiang:2021jbg} (see also the more recent study \cite{Brizio:2024doe}) as we discuss later. { In the following, for brevity, we will use the denomination ``$\TTb$ deformation'' for both the original deformation of \cite{Smirnov:2016lqw,Cavaglia:2016oda} and for the generalised, higher-spin versions, confident that the context-awareness of the reader will avoid potential confusions.}
 
 It has been know for a long time that in the presence of a boundary, a new set of functions $R_a(\theta)$ come to play a prominent role in the formulation of the scattering theory of the model. The $S$-matrix remains unchanged but scattering processes off the boundary now need to be accounted for, while retaining integrability. In this context, the function $R_a(\theta)$ is the reflection amplitude off the boundary. Unitarity and crossing relations lead to the constraints \cite{Ghoshal:1993tm}:
 \beq
 R_a(\theta)R_a(-\theta)=1 \quad \mathrm{and} \quad R_a(\theta) R_{\bar{a}}(\theta+i\pi)=S_{aa}(2\theta)\,, \label{uc}
 \eeq 
 where $\bar{a}$ is the particle conjugate to $a$. As we can see, reflection amplitudes are related to the scattering phase. Thus, when this changes, like in the presence of irrelevant perturbations, we expect $R_a(\theta)$ to also change.  
In the presence of stable bound states \cite{Cherednik:1984vvp,Sklyanin:1988yz} there are additional requirements for the functions $R_a(\theta)$ in the form of boundary bootstrap equations. These take the form:
 \beq 
R_a(\theta+i\eta_{ac}^b) R_b(\theta+i \eta_{bc}^a) S_{ab}(\theta+i\eta_{bc}^a+i\eta_{ac}^b)=R_c(\theta)\,,
 \eeq 
 where $\eta_{ab}^c$ are values related to the position of the poles of the scattering matrix. If  the $S$-matrix $S_{ab}(\theta)$ has a pole at $\theta=i u_{ab}^c$ corresponding to the formation of the bound state $c$ in the process $a+b \mapsto c$, then $\eta_{ab}^c:=\pi-u_{ab}^c$. Systematic solutions to these equations have been famously constructed for affine Toda field theories \cite{Corrigan:1994np,ATFTB,MeandA}.

This program can be further extended by considering the possibility of a ``dynamical" boundary, namely a boundary which can be excited to a different state by particle collision. This is associated with a pole of the reflection amplitudes themselves. This possibility was first put forward in \cite{fring}. Notably, it is fully compatible with integrability. 
If we label the type of boundary by capital letters, then we can see this as the process $a+A \mapsto B$. In this case, boundary reflection amplitudes acquire an extra index $R_a^A(\theta)$ and there are additional boundary bootstrap equations
\beq 
R_a^A(\theta)=R_a^B(\theta) S_{ab}(\theta+i\eta_{aA}^B)S_{ab}(\theta-i\eta_{aA}^B)\,,
\eeq 
where $\theta=i\eta_{aA}^B$ is a pole of the amplitude $R_a^A(\theta)$. The solutions to these equations in Toda field theories where studied in great detail in \cite{Riva:2001pb}.

Once solutions to these equations have been found, they can be employed as input data in the study of the thermodynamic properties of massive boundary IQFTs as done in \cite{LeClair:1995uf,Bajnok:2007ep}. They can also be employed in the context of computing correlation functions and their building blocks (form factors). This may be done either by employing the boundary state as proposed in \cite{Ghoshal:1993tm} or by developing a form factor program for boundary IQFTs, as done in \cite{BFF} and employed for example in \cite{MeA2,MeSG,Kormos:2007qx,Takacs:2008zz, Lencses:2011ab,Bajnok:2015iwa,Hollo:2014vpa}. 

\medskip

This paper is organised as follows: In Section \ref{1} we review the construction of deformed reflection amplitudes, starting from (\ref{s}). This overlaps with the work \cite{Jiang:2021jbg} but is presented here in the more restrictive setting of IQFT.  In Section \ref{2} we review the boundary form factor program, focusing only on one-particle form factors, particularly the so-called minimal part. In Section \ref{4} we introduce the set of reflection amplitudes of the sinh-Gordon theory and discuss the special case of Dirichlet boundary conditions. In Section \ref{min5} we show that the minimal form factor admits a representation of the $\TTb$ type. This representation is functionally similar to the results of \cite{MFF}. In Section \ref{6} we discuss the extension of our construction to more general boundary conditions. We conclude in Section \ref{3}.

\section{Reflection Amplitudes and Irrelevant Perturbations}
\label{1}
Consider for simplicity a theory with no bound states. %and a single particle content. We can from now on, drop all particle indices so that the $S$-matrix is $S(\theta)$, the deformed $S$-matrix is $S^{\bal}(\theta)$ and the CDD factor is $\Phi^{\bal}(\theta)$.
Let $R_a(\theta)$ be a reflection amplitude which preserves integrability and has no pole leading to excited boundary states. 
%Several interesting examples of such a solution are known. For instance, the Ising field theory allows for an interesting family of distinct boundary conditions leading to various solutions for $R(\theta)$ parametrized by the boundary magnetic field \cite{Ghoshal:1993tm}. As we shall see later, the sinh-Gordon model also admits a family of solutions, depending on two parameters \cite{CorTao}. 
In this case the only relevant equations for $R_a(\theta)$ are (\ref{uc}). We will now promote $R_a(\theta)$ to $R_a^{\bal}(\theta)$ to denote the deformed solution to equations (\ref{uc}) corresponding to the deformed $S$-matrix (\ref{s}). We expect that 
\beq 
R_a^{\bal}(\theta)= R_a(\theta) \Lambda_a^{\bal}(\theta)\,,
\label{5}
\eeq 
for some function $\Lambda_a^{\bal}(\theta)$ which satisfies
\beq 
\Lambda_a^{\bal}(\theta)\Lambda_a^{\bal}(-\theta)=1 \qquad \mathrm{and} \qquad \Lambda_a^{\bal}(\theta)\Lambda_a^{\bal}(\theta+i\pi)=\Phi_{aa}^{\bal}(2\theta)\,.
\eeq 
It is very easy to see that these equations are solved by 
\beq 
\Lambda_a^{\bal}(\theta)=\sqrt{\Phi_{aa}^{\bal}(2\theta)}\,.
\label{eq:Lambda_refl_ampl}
\eeq 
This is the standard type of solution, namely a $2\pi i$ periodic, odd function of $2\theta$ and it agrees with the boundary scattering factor found in \cite{Jiang:2021jbg}.
This gives the universal change of the reflection amplitudes in boundary IQFTs after a $\TTb$ perturbation.
The solutions for $\Lambda_a^{\bal}(\theta)$ can however be more general than this. While the factor $\sqrt{\Phi_{aa}^{\bal}(2\theta)}$ needs to be there,  any function of the type $\sinh(k \theta)$ with $k$ odd can be added to the exponent, providing a new solution to (\ref{uc}). In general we have 
\beq 
\Lambda_a^{\bal}(\theta)= \sqrt{\Phi^{\bal}_{aa}(2\theta)} \exp\left[{-i \sum_{k \in \mathbb{Z}} \gamma_k m^{2(2k+1)} \sinh((2k+1)\theta)}\right]\,.
\label{8}
\eeq 
{ Therefore we obtain multiple possible deformations for the same reflection amplitude. The presence of this type of ambiguities or CDD factors is also common when computing two-body scattering amplitudes using the bootstrap program. This is because the $S$-matrix bootstrap equations generically have many distinct solutions. However, in most cases, the solution can be narrowed down by utilising additional information about the theory, such as its semiclassical spectrum or UV limit. Typically, we can then identify a unique solution. In the context of $\TTb$-like deformations, the $S$-matrix deformation can be uniquely defined for example by employing the $JT$-like gravity formulation \cite{Dubovsky:2012wk}. Once an $S$-matrix is fixed, multiple solutions for the reflection amplitudes are still expected since, in general, there are several integrable boundary conditions allowed for one single scattering amplitude\footnote{Many interesting examples are known. The simplest cases are the Ising field theory, where a family of distinct boundary conditions exist parametrised by the boundary magnetic field \cite{Ghoshal:1993tm}. Similarly, the sinh-Gordon model admits a two-parameter family of solutions, as found in \cite{CorTao}. We will discuss these two models  in Sections \ref{4}-\ref{6}.}. Furthermore, as shown in \cite{MeandA} for affine Toda field theories, reflection amplitudes associated to the same $S$-matrix but distinct boundary conditions can be related to each other by simple multiplication with hyperbolic function blocks, that is, once more CDD factors. It is this property that the exponential in (\ref{8}) represents. From here onwards, we will take the simplest solution $\gamma_i=0$. } 

The question of how the reflection amplitudes are deformed under irrelevant perturbations has already been discussed in the literature a few years ago \cite{Jiang:2021jbg} and then employed to develop a generalised boundary thermodynamic Bethe ansatz. { In their work, the ambiguity (\ref{8}) is fixed by construction}. Here we proceed instead to discuss the form factor program in the presence of boundaries. 

\section{Boundary Form Factor Program and Minimal Form Factor}
\label{2}
The boundary form factor program was proposed in \cite{BFF} and subsequently applied to several models and fields \cite{MeA2,MeSG,Kormos:2007qx,Takacs:2008zz, Lencses:2011ab,Bajnok:2015iwa,Hollo:2014vpa}. The main idea of the program is to address the computation of correlation functions in the presence of a boundary. There are two viewpoints we may take. If the boundary is located at the origin of time it can be represented by a boundary state in the Ghoshal-Zamolodchikov sense \cite{Gross:2019ach}. In this case matrix elements of local fields may be computed in terms of the matrix elements obtained in the absence of a boundary, assuming these are known via the standard form factor program \cite{Karowski:1978vz,smirnov1992form}. This can be achieved by expanding the boundary state in terms of bilinears of the Zamolodchikov-Fadeev algebra. It is also possible to think of the boundary as located in space, say at the origin. In this case, one can derive a set of modified form factor equations for the matrix elements of local fields which now must take into account scattering processes off the boundary. These equations were presented in \cite{BFF}.  { In this paper we focus mainly on the fundamental building blocks for higher particle form factors, that is the one- and two-particle form factors of a local field $\mathcal{O}$. }
\subsection{One-Particle Form Factors}
{ The one-particle form factor equations are simply:}
\beq 
F^{\mathcal{O}}_a(\theta)=R_a(\theta)F^\mathcal{O}_a(-\theta)\qquad \mathrm{and} \qquad F^{\mathcal{O}}_a(\theta)=R_a(i\pi-\theta)F^\mathcal{O}_a(2\pi i-\theta)\,,
\label{1min}
\eeq 
where
\beq 
F^{\mathcal{O}}_a(\theta):= \bra 0|{\mathcal{O}}(0)|\theta \ket_a\,,
\eeq 
with $|\theta\ket_a$ an in-state containing a single particle of species $a$ and $|0\ket$ the vacuum state. Due to breaking of translation invariance, the one-particle form factor is rapidity dependent, even for spinless fields. Thus, it is the simplest non-trivial form factor that may be computed and a building block for higher particle form factors. 

{ Let us denote by $r_a^{\rm min}(\theta)$ a minimal solution to the equations (\ref{1min}). The solution procedure} was presented in \cite{BFF} and follows the usual construction. Starting with an integral representation for $R_a(\theta)$, { a corresponding integral representation of $r_a^{\rm min}(\theta)$} can be found. The minimal solution to the equations, without poles in the physical strip, is entirely determined by $R_a(\theta)$ whereas the pole structure of $F_a^\mathcal{O}(\theta)$ is related to the operator $\mathcal{O}$. { In \cite{BFF} it was shown that the full solution to the form factor equations must take the form:
\beq 
F_a^\mathcal{O}(\theta)=r_a^{\rm min}(\theta) Q^{\mathcal{O}}_a(y)\,\quad \mathrm{with}\quad y=2\cosh\theta,
\eeq 
with $Q_a^{\mathcal{O}}(y)$ and operator-dependent function.
}

Let $r_a^{\rm{min}}(\theta)$ be a minimal solution of (\ref{1min}). What would be its deformed version in the presence of a generalised $\TTb$ perturbation? Following 
\cite{longpaper,ourentropy,theirentropy} we observe that the equations (\ref{1min})  are linear and factorised. Therefore, we expect that the modified minimal form factor to be of the form
\beq 
r_a^{\rm{min}}(\theta;\bal)= r_a^{\rm{min}}(\theta) \varphi^{\bal}_a(\theta)\,,
\label{even}
\eeq
with 
\beq 
\varphi^{\bal}_a(\theta)=\Lambda_{a}^{\bal}(\theta)\varphi^{\bal}_a(-\theta)
\qquad \mathrm{and} \qquad \Lambda_{a}^{\bal}(\theta)\varphi^{\bal}_a(\theta)=\varphi^{\bal}_a(2\pi i -\theta)\,,
\label{11min}
\eeq 
which is solved by
\beqa 
\log\varphi_{a}^{\bal}(\theta)=\frac{2\theta-i\pi}{2\pi}i\log\Lambda_a^{\bal}(\theta)=\frac{2\theta-i\pi}{4\pi}i \log\Phi_{aa}^{\bal}(2\theta) \,.
\label{phi}
\eeqa 
However, there is a larger family of solutions. Indeed, the exponent above, can be modified by a sum of $\cosh(k\theta)$ functions with $k \in \mathbb{Z}$, and still satisfy all requirements. We therefore find that the most general minimal solution to \eqref{1min} is
\beq 
r_a^{\rm min}(\theta;\bal,\bel)=r_a^{\rm{min}}(\theta) \varphi^{\bal}_a(\theta) C_a^{\bel}(\theta)\,,
\label{14}
\eeq 
with $\varphi^{\bal}_a(\theta) $ given by \eqref{phi} and
\beq 
\log C_a^{\bel}(\theta)= \sum_{s \in \mathcal{S}'} \beta_s m_a^{2s} \cosh(s\theta)\,.
\label{15}
\eeq 
Thus, the solution is parametrised by parameters $\bal$, which are determined by the deformation of the $S$-matrix, and $\bel$ which can in principle be freely chosen.

The presence of free parameters in the minimal form factor is an issue that we also encountered when considering theories without boundaries \cite{PRL,longpaper,ourentropy,theirentropy} and whose meaning, for the time being, is not fully understood. Traditionally, we would expect the minimal form factor to be entirely fixed by analyticity and asymptotics requirements. Indeed, this is the case for standard IQFTs, as we shall see in the next section. However, for models perturbed by a finite number of irrelevant perturbations finding the natural choice of parameters $\bel$ remains difficult. Progress in this direction will be reported soon \cite{inprep}.

{ \subsection{Two-Particle Form Factors} 
The boundary form factor equations for the two-particle form factors take the form:
\beq 
F^{\mathcal{O}}_{ab}(\theta_1,\theta_2)=S_{ab}(\theta_1-\theta_2)F^\mathcal{O}_{ba}(\theta_2,\theta_1)\,,\qquad F^{\mathcal{O}}_{ab}(\theta_1,\theta_2)=R_b(\theta_2)F^\mathcal{O}_{ab}(\theta_1,-\theta_2)
\label{2min}
\eeq 
and 
\beq 
F_{ab}^{\mathcal{O}}(i\pi+\theta_1,\theta_2)=R_a(-\theta_1)F^\mathcal{O}_{ab}(\pi i-\theta_1,\theta_2)
\eeq 
where 
\beq 
F_{ab}^{\mathcal{O}}(\theta_1,\theta_2):=\bra 0|\mathcal{O}(0)|\theta_1 \theta_2\ket_{ab}\,,
\eeq 
with $|\theta_1 \theta_2\ket_{ab}$ and in-state containing two particles of species $a$ and $b$ and rapidities $\theta_1,\theta_2$. Note that, contrary to the bulk case, the two-particle form factor is no longer just a function of rapidity differences. 
In \cite{BFF} a strategy was presented to find a minimal solution to these three equations. It was found that the two-particle form factor must generally have the form:
\beq 
F_{ab}^{\mathcal{O}}(\theta_1,\theta_2)= r^{\rm min}_a(\theta_1)r^{\min}_b(\theta_2) f^{\rm min}_{ab}(\theta_1-\theta_2)f^{\rm min}_{ab}(\theta_1+\theta_2) Q_{ab}^{\mathcal{O}}(y_1,y_2)\,,
\label{2par}
\eeq 
where $f^{\rm min}_{ab}(\theta)$ is the two-particle minimal form factor in the bulk, that is the minimal solution to the equations
\beq 
f^{\rm min}_{ab}(\theta)=S_{ab}(\theta)f^{\rm min}_{ab}(-\theta)=f^{\rm min}_{ab}(2\pi i-\theta)\,,
\eeq 
and $Q_{ab}^{\mathcal{O}}(y_1,y_2)$ is an operator-dependent function, which would includes any poles present in the form factor and must be a function of the variables $y_i:=2\cosh\theta_i$. We can say that there is a boundary two-particle minimal form factor which we can define as the universal part of (\ref{2par})
\beq 
r_{ab}^{\rm min}(\theta_1,\theta_2):=r^{\rm min}_a(\theta_1)r^{\min}_b(\theta_2) f^{\rm min}_{ab}(\theta_1-\theta_2)f^{\rm min}_{ab}(\theta_1+\theta_2)\,.
\eeq 
In this work, we assume that the form factor equations remain unchanged in the presence of irrelevant perturbations\footnote{Note that this is a non-trivial assumption which we have also made in our previous works \cite{PRL,longpaper}. We plan to investigate this point further in future works \cite{inprep}.}. It follows then that for the deformed theory, the function above should just be lifted to:
\beq 
r_{ab}^{\rm min}(\theta_1,\theta_2;\bal,\bel,\hat{\bel}):=r^{\rm min}_a(\theta_1;\bal,\bel)r^{\min}_b(\theta_2;\bal,\bel) f^{\rm min}_{ab}(\theta_1-\theta_2;\bal,\hat{\bel})f^{\rm min}_{ab}(\theta_1+\theta_2;\bal,\hat{\bel})\,,
\eeq 
where the minimal one-particle form factors are those found above (\ref{14}) and the deformed two-particle minimal form factor in the bulk, was found in \cite{PRL,longpaper}
\beq
f^{\rm min}_{ab}(\theta;\bal,\hat{\bel})=f^{\rm min}_{ab}(\theta) \varphi_{ab}^{\bal} (\theta)C^{\hat{\bel}}_{ab}(\theta)\,,
\eeq 
with $f^{\rm min}_{ab}(\theta)$ the underformed two-particle minimal form factor in the bulk, and 
\beq 
\log(\varphi^{\bal}_{ab}(\theta))=\frac{ \theta-i\pi}{2\pi} i \log(\Phi^{\bal}_{ab}(\theta))\,,\quad \mathrm{and} \quad \log(C^{\hat{\bel}}_{ab}(\theta))=\sum_{s\in \mathcal{S}'} \hat{\beta}_s m_a^{s} m_b^s \cosh(s\theta)\,,
\label{mini}
\eeq
where $\hat{\bel}$ are arbitrary parameters. 
\subsection{Higher Particle Form Factors and Correlation Functions}
The construction of subsections 3.1 and 3.2 can be continued to higher particle form factors by starting with the natural ansatz that the solutions above suggest, namely
\beq 
F_{a_1 \ldots a_n}^{\mathcal{O}}(\theta_1,\ldots,\theta_n)= Q_{a_1\ldots a_n}^{\mathcal{O}}(y_1,\ldots,y_n)\prod_{j=1}^n r^{\rm min}_{a_i}(\theta) \prod_{1\leq i<j \leq n} f_{a_i a_j}^{\rm min}(\theta_1+\theta_2) f_{a_i a_j}^{\rm min}(\theta_1-\theta_2)\,,
\eeq 
once more, this can be easily extended to the $\TTb$ perturbed case by introducing dependencies on the parameters $\bal$, $\bel$ and $\hat{\bel}$. As discussed also in \cite{BFF}, in the unperturbed case, the $Q_{a_1\ldots a_n}^{\mathcal{O}}(y_1,\ldots,y_n)$ are rational functions which incorporate the pole structure, including both bulk and boundary kinematic poles. The latter give rise to denominators involving products of $y_i+y_j$ with $i<j$ and/or products of just $y_j$, respectively. We will leave the systematic study of the solutions to these equations in the $\TTb$-perturbed case for future work. 

We would like to end this section by making a general observation about correlation functions. It is well known that the form factors are building blocks for correlation functions. The minimal form factors presented above have a distinct feature that will play a key role in the asymptotics of correlators. Consider for simplicity the case $\bel=\hat{\bel}=\bol$ and one single non-vanishing $\alpha_s$, say $\alpha:=\alpha_1$, the $\TTb$ perturbation. We have that 
\beq 
|r_a^{\rm min}(\theta;\alpha,\bol)|^2=|r_a^{\rm min}(\theta)|^2 |\varphi_a^{\alpha}(\theta)|^2=|r_a^{\rm min}(\theta)|^2 e^{\frac{2 m_a^s m_b^s \alpha}{\pi}  \theta \sinh(2 s\theta)}\,,
\eeq
and, similarly, 
\begin{equation}
	\begin{split}
		|r_{ab}^{\rm min}&(\theta_1,\theta_2;\alpha,\bol,\bol)|^2 =|r_a^{\rm min}(\theta)|^2 |r_b^{\rm min}(\theta)|^2 ||f^{\rm min}_{ab}(\theta_1-\theta_2)|^2  |f^{\rm min}_{ab}(\theta_1+\theta_2)|^2 \\
		&\times \varphi_a^{\alpha}(\theta)|^2|\varphi_b^{\alpha}(\theta)|^2 |\varphi_{ab}^{\alpha} (\theta_1-\theta_2)|^2 |\varphi_{ab}^{\alpha} (\theta_1+\theta_2)|^2\\
		&= |r_a^{\rm min}(\theta)|^2 |r_b^{\rm min}(\theta)|^2 ||f^{\rm min}_{ab}(\theta_1-\theta_2)|^2  |f^{\rm min}_{ab}(\theta_1+\theta_2)|^2 \\
		&\times \exp\left[\frac{2\alpha }{\pi} \Big(m_a^{2s} \theta_1 \sinh(2s\theta_1) +  m_b^{2s} \theta_2 \sinh(2s\theta_2)\Big) \right. \\
		&+ \left.\frac{\alpha m_a^s m_b^s}{\pi} \Big((\theta_1+\theta_2)\sinh(s(\theta_1+\theta_2))+  (\theta_1-\theta_2) \sinh(s(\theta_1-\theta_2))\Big)\right]\,.
	\end{split}
\end{equation}
These quantities will enter the form factor expansion of a typical two-point function in the ground state. What is important is that these are functions that are rapidly increasing/decreasing in the rapidity variables for $\alpha$ positive/negative. In the $\alpha>0$ case this means that any form factor expansion of the correlation function will be divergent, whereas for $\alpha<0$ it will be very rapidly convergent. Indeed, convergence is so strong for $\alpha<0$ that higher particle form factors will provide negligible contributions to the form factor expansion. This behaviour has also been found in the bulk case 
 \cite{PRL,longpaper} and is consistent with the observation that there is a stark difference between the regimes of positive and negative coupling, as found in the TBA analysis \cite{Cavaglia:2016oda,Jiang:2021jbg}. This behaviour is robust under the reintroduction of the $\bel$ and $\hat{\bel}$ parameters, as long as their number is finite. As we shall see in the following section, when the number of such parameter is infinite, the asymptotic properties of the minimal form factor can be radically different. 
 \medskip
 
In the next two Sections we will focus our attention on a known integrable quantum field theory (the sinh-Gordon model) and demonstrate that its boundary one-particle minimal form factors admit a new representation which consists of blocks of the form (\ref{14}). In this representation, the ``unperturbed" minimal form factor is the minimal form factor of the Ising field theory with specific boundary conditions. The idea that the sinh-Gordon theory (with and without boundaries) may be seen as a perturbation of the Ising field theory was also exploited in \cite{MFF} to find a new representation of the bulk form factor. This result is significant for two main reasons: it confirms that the structure of the deformations in (\ref{14}) is widespread in IQFT and it provides a more numerically efficient representation for a function which plays a key role in evaluation of correlation functions.
 }

\section{The sinh-Gordon Model with Dirichlet Boundary Conditions}
\label{4}
The sinh-Gordon model with Dirichlet boundary conditions was one of the examples considered in \cite{BFF} and later in \cite{MeSG}. This particular choice of boundary conditions has the advantage that the minimal form factor $r_a^{\rm min}(\theta)$ coincides with the one-particle form factor, that is, there are no additional poles to be included, which makes calculations particularly simple. In this case non-vanishing form factors associated with odd particle numbers can be identified as corresponding to the operator $\partial_x \phi$, where $\phi$ is the sinh-Gordon field. More generally, as reported in \cite{CorTao}, there is a two-parameter family of solutions for the reflection amplitudes of the sinh-Gordon model. They can be written in terms of fundamental blocks $(x)_\theta$, $[x]_\theta$ as
\beq
R(\theta,E,F)=\left(\frac{1}{2}\right)_\theta\left(\frac{2+B}{4}\right)_\theta\left(1-\frac{B}{4}\right)_\theta\left[\frac{E-1}{2}\right]_\theta\left[\frac{F-1}{2}\right]_\theta\,,
\label{GenR}
\eeq 
where\footnote{Note that there is a minus sign difference between the definitions of $[x]_\theta$ in \cite{BFF} and \cite{MFF}. Here we are using the same definitions as in \cite{MFF}. Compared to \cite{CorTao} the blocks $(x)_\theta$ differ by a factor $1/2$ in the definition of $x$.}
\beq
(x)_\theta:=\frac{\sinh\frac{1}{2}(\theta+i\pi x)}{\sinh\frac{1}{2}(\theta-i\pi x)}\,,\qquad [x]_\theta=-(x)_\theta (1-x)_\theta=\frac{\tanh\frac{1}{2}(\theta+i\pi x)}{\tanh\frac{1}{2}(\theta-i\pi x)}\,.
\eeq
The sinh-Gordon two-body scattering matrix is simply \beq 
S(\theta)=\left[-\frac{B}{2}\right]_\theta=\frac{\sinh\theta-i \sin\frac{\pi B}{2}}{\sinh\theta+i \sin\frac{\pi B}{2}}\,,
\eeq 
with $B\in [0,2]$ a coupling constant \cite{SSG,SSG2,SSG3}. The simplest version of (\ref{GenR}) is obtained by
removing the $F$-dependent factor and by setting $E=0$, while introducing an overall minus sign (this is due to the particular definition of our $[x]_\theta$ symbol, as explained in footnote 2).  This corresponds to Dirichlet boundary conditions that fix the boundary field to 0. In that special case, the amplitude (\ref{GenR}) reduces to
\beq 
R(\theta)=-\left(-\frac{1}{2}\right)_\theta\left(\frac{2+B}{4}\right)_\theta\left(1-\frac{B}{4}\right)_\theta\,.
\label{thisR}
\eeq  
%\FS{I think there is a minus in front of the blocks. In fact if you send B=0 you get a -1 from the blocks.}
This choice also cancels out the pole of the reflection amplitude at $\theta=\frac{i\pi}{2}$ that is present in (\ref{GenR}) due to the block $(\frac{1}{2})_\theta$. A special property of this amplitude is that for $B=0$ it reduces to $R(\theta)=1$ which corresponds to a free boson solution (the sinh-Gordon $S$-matrix reduces to $1$ for $B=0$). The minimal form factor solution corresponding to this free boson case is proportional to $\sinh\theta$. In \cite{BFF}, the minimal form factor solution corresponding to (\ref{thisR}) was given as
\beq 
r^{\rm min}(\theta)=\frac{\sinh\theta}{i+\sinh\theta} u(\vartheta,B)\,,
\label{for}
\eeq 
with 
\beq 
u(\vartheta,B)=\exp\left[-2\intop_0^\infty \frac{dx}{x} \left(\cos\frac{\vartheta x}{\pi} -1\right)\frac{\cosh\frac{x}{2}}{\sinh^2 x} \left(\sinh\frac{xB}{4}+\sinh\left(1-\frac{B}{2}\right)\frac{x}{\
2} +\sinh\frac{x}{2} \right)\right]\,,
\label{mffSG}
\eeq 
where $\vartheta=\frac{i\pi}{2}-\theta$. The  normalisation is chosen so that $u(0,B)=1$. We have that 
\beq
u(\vartheta,0)=\exp{\left[-2\intop_0^\infty \frac{dx}{x} \left(\cos\frac{\vartheta x}{\pi}-1\right) \frac{\cosh\frac{x}{2}}{\sinh^2 x} \left(2\sinh\frac{x}{2} \right)\right]}=-\frac{i}{2}(i+\sinh\theta)\,,
\eeq 
so that for $B=0$ we recover the free boson solution $r^{\rm min}(\theta)=-\frac{i}{2}\sinh\theta$. For our purposes however, it is interesting to emphasise the connection with free fermions instead. The sinh-Gordon model is a fermionic theory, in the sense that $S(\theta=0)=-1$ and if we factor out this $-1$ from the $S$-matrix $[-B/2]_\theta$ what remains can be seen as a CDD factor. This means that the sinh-Gordon $S$-matrix is of the type (\ref{s}) with a CDD factor {given by} a sum over all odd integers and coefficients $m^{2s} \alpha_s$ which are functions of $B$. The precise formulae were discussed in \cite{MFF}. {Hence}, according to our derivation in Section \ref{2}, the minimal form factor (\ref{mffSG}) should also admit a representation of the type (\ref{14}) with (\ref{phi}). We will now show that this is indeed the case.
\section{A New Minimal Form Factor Representation}
\label{min5}
Let
\beq 
\omega(\vartheta, B):=\log u(\vartheta,B)\,.
\eeq 
then, the derivative w.r.t. $\vartheta$ is,
\beqa 
\omega'(\vartheta,B)= h(\vartheta,B)+h(\vartheta,2-B)+g(\vartheta)\,,
\eeqa 
with 
\beq 
h(\vartheta,B)=\frac{2}{\pi} \intop_0^\infty dx  \frac{\sin\frac{\vartheta x}{\pi}\cosh\frac{x}{2} \sinh\frac{xB}{4}}{\sinh^2 x}\qquad \mathrm{and} \qquad  g(\vartheta)= \frac{1}{\pi} \intop_0^\infty dx \frac{\sin\frac{\vartheta x}{\pi} }{\sinh x}\,.
\label{hdef}
\eeq 
We have that $g(\vartheta)$ can be easily integrated to $g(\vartheta)=\frac{1}{2}\tanh\frac{\vartheta}{2}$, while $h(\vartheta,B)$  can be computed using contour integration, along the same lines of the computations presented in \cite{MFF}. For example, we have the integral
\beq 
I(a,b)=\intop_{-\infty}^\infty  dx \frac{e^{(b+ i a) x}}{\sinh^2 x}= i \pi (b+ i a) \frac{1+ e^{i\pi (b+ i a)}}{1-e^{i\pi (b+ i a)}}\,,
\label{Iab}
\eeq 
from where it follows
\beqa 
h(\vartheta,B)&=& \frac{1}{8\pi i} \left[I(\frac{\vartheta}{\pi},\frac{2+B}{4})+I(\frac{\vartheta}{\pi},\frac{-2+B}{4})-I(\frac{\vartheta}{\pi},\frac{2-B}{4})-I(\frac{\vartheta}{\pi},-\frac{2+B}{4})\right.\nonumber\\
& -&\left. I(-\frac{\vartheta}{\pi},\frac{2+B}{4})-I(-\frac{\vartheta}{\pi},\frac{-2+B}{4})+I(-\frac{\vartheta}{\pi},\frac{2-B}{4})+I(-\frac{\vartheta}{\pi},-\frac{2+B}{4})\right] = \nonumber\\
&=& \frac{4\vartheta \sin\frac{\pi B}{2}-4\pi \sin\frac{\pi B}{4}\sinh\vartheta+ \pi B \sinh(2\vartheta)}{4\pi (\cosh(2\vartheta)+\cos\frac{\pi B}{2})}.
\label{hint}
\eeqa 

Integrating gives 
\beqa
&& \omega(\vartheta,B)= \log a(B)+\log \cosh\frac{\vartheta}{2}-\frac{i\vartheta}{2\pi}\log \left[\frac{\sin\frac{B\pi}{2}+i\sinh(2\vartheta)}{\sin\frac{B\pi}{2}-i\sinh(2\vartheta)}\right]\nonumber\\
&& +\frac{1}{2}\log\left[\cos\frac{B\pi}{4}+\cosh\vartheta\right]+\frac{1}{4}\log\left[\frac{\cosh\vartheta+ \sin\frac{B\pi}{4}}{\cosh\vartheta-\sin\frac{B\pi}{4}}\right]-\frac{B}{8}\log\left[\frac{\cosh(2\vartheta)-\cos\frac{\pi B}{4}}{\cosh(2\vartheta)+\cos\frac{\pi B}{4}}\right] \nonumber\\
&& +
\frac{i}{4\pi}\left( {\rm Li}_2(-ie^{-\vartheta-\frac{i\pi B}{4}})+{\rm Li}_2(ie^{-\vartheta-\frac{i\pi B}{4}})-{\rm Li}_2(-ie^{-\vartheta+\frac{i\pi B}{4}})-{\rm Li}_2(ie^{-\vartheta+\frac{i\pi B}{4}})+\vartheta \mapsto -\vartheta\right)\nonumber\\
&& -\frac{i}{4\pi}\left({\rm Li}_2(-e^{-\vartheta-\frac{i\pi B}{4}})+{\rm Li}_2(e^{-\vartheta-\frac{i\pi B}{4}})-{\rm Li}_2(-e^{-\vartheta+\frac{i\pi B}{4}})-{\rm Li}_2(e^{-\vartheta+\frac{i\pi B}{4}})+\vartheta \mapsto -\vartheta\right)\,. \label{mainfor}
\eeqa 
Here $\log a(B)$ is an integration constant which can be fixed by asymptotic requirements. Requiring that $\omega(0,B)=0$ and after some simplifications, we obtain
\beqa
\log a(B)&=& -\frac{1}{2}\log\left(\frac{\sin\frac{\pi(2+B)}{8}}{\sin\frac{\pi(2-B)}{8}}\right)-\log\left(2\cos\frac{B\pi}{8}\right)+\frac{B}{4}\log\tan\frac{B\pi}{4}\nonumber\\
&& + \frac{i}{2\pi}\left({\rm Li}_2(-e^{\frac{i\pi B}{2}})-{\rm Li}_2(e^{\frac{i\pi B}{2}})\right)-\frac{i\pi (B-1)}{8}\,.
\eeqa 
It is interesting to consider the various contributions to (\ref{mainfor}):
\begin{itemize}
\item The contribution 
\beq \log \cosh\frac{\vartheta}{2}=\log \cos\frac{1}{2}\left(\frac{i\pi}{2}-\theta
\right)\,,
\eeq 
is such that when taking exponential of $\omega(\vartheta,B)$ it gives a factor $\cosh\frac{1}{2}\left(\frac{i\pi}{2}-\theta
\right)$ in the minimal form factor that combines with the prefactor $\frac{\sinh\theta}{\sinh\theta+ i}$ in (\ref{for}) to give 
\beq 
r_{\rm fixed}(\theta)=-\frac{i}{2}\frac{\sinh\theta}{\cosh\frac{1}{2}\left(\frac{i\pi}{2}-\theta
\right)}\,,
\label{rIsing}
\eeq 
which is the minimal form factor corresponding to the Ising model with reflection amplitude
\beq 
R^{\bol}(\theta)=\left(-\frac{1}{2}\right)_\theta\,.
\eeq 
This is known as the fixed boundary condition of the Ising model, and corresponds to the limit of infinite boundary magnetic field, as discussed in \cite{Ghoshal:1993tm,BFF} (see also Section 6 for further discussion).
\item The contribution 
\beq 
-\frac{i\vartheta}{2\pi}\log \left[\frac{\sin\frac{B\pi}{2}+i\sinh(2\vartheta)}{\sin\frac{B\pi}{2}-i\sinh(2\vartheta)}\right]=\frac{2\theta-i\pi}{4\pi}i\log(-S(2\theta))=\frac{2\theta-i\pi}{4\pi}i\log\Phi^{\rm shG}(2\theta) \,,
\eeq
where $\Phi^{\rm shG}(\theta)$ is minus the scattering matrix of the sinh-Gordon model, which can be seen as a CDD factor. Hence, sinh-Gordon emerges as a perturbation of the Ising field theory. 
\item The remaining terms in (\ref{mainfor}) add up to an even function of $\theta$ which admits a formal expansion as a sum of $\cosh(s\theta)$ functions with $s$ integer, both odd and even. This is similar to the computations presented in the Appendix of \cite{MFF}.  
\end{itemize}
In summary, the minimal form factor $r^{\rm min}(\theta)$ introduced in (\ref{for}) can be rewritten as
\beq 
r^{\rm min}(\theta)= a(B) r_{\rm fixed}(\theta) e^{\frac{2\theta-i\pi}{4\pi} i \log (-S(2\theta))} C^{\bel}(\theta)\,,
\label{solutionShG}
\eeq 
with $r_{\rm fixed}(\theta)$ given by (\ref{rIsing}), $S(\theta)$ the sinh-Gordon $S$-matrix and $C^{\bel}(\theta)$ given by
\beqa
&& \log C^{\bel}(\theta)\nonumber\\
&& = \frac{1}{2}\log\left[\cos\frac{B\pi}{4}+\cosh\vartheta\right]+\frac{1}{4}\log\left[\frac{\cosh\vartheta+ \sin\frac{B\pi}{4}}{\cosh\vartheta-\sin\frac{B\pi}{4}}\right]-\frac{B}{8}\log\left[\frac{\cosh(2\vartheta)-\cos\frac{\pi B}{4}}{\cosh(2\vartheta)+\cos\frac{\pi B}{4}}\right] \nonumber\\
&& +
\frac{i}{4\pi}\left( {\rm Li}_2(-ie^{-\vartheta-\frac{i\pi B}{4}})+{\rm Li}_2(ie^{-\vartheta-\frac{i\pi B}{4}})-{\rm Li}_2(-ie^{-\vartheta+\frac{i\pi B}{4}})-{\rm Li}_2(ie^{-\vartheta+\frac{i\pi B}{4}})+\vartheta \mapsto -\vartheta\right)\nonumber\\
&& -\frac{i}{4\pi}\left({\rm Li}_2(-e^{-\vartheta-\frac{i\pi B}{4}})+{\rm Li}_2(e^{-\vartheta-\frac{i\pi B}{4}})-{\rm Li}_2(-e^{-\vartheta+\frac{i\pi B}{4}})-{\rm Li}_2(e^{-\vartheta+\frac{i\pi B}{4}})+\vartheta \mapsto -\vartheta\right)\,, 
\eeqa 
with $\vartheta=\frac{i\pi}{2}-\theta$ as before. As indicated by \eqref{15}, this function admits a formal expansion in terms of $\cosh(\ell\theta)$ functions. We can show that
\begin{equation}
    \begin{split}
        \log C^{\boldsymbol{\beta}}(\theta) & = - \frac{1}{2}\log 2 - \sum_{\ell=1}^\infty \frac{1}{\ell}\cos\left(\frac{B}{4}\ell \pi\right)\cosh(\ell \theta)  \\
        & - \sum_{\ell=0}^\infty \frac{(-1)^\ell}{2\ell + 1} \sin\left(\frac{B}{4}(2\ell+1)\pi\right) \cosh((2\ell+1)\theta)  \\
        & + \frac{B}{2} \sum_{\ell=0}^\infty \frac{1}{2\ell+1} \cos\left(\frac{B}{4}(2\ell+1)\pi\right) \cosh(2(2\ell+1)\theta)  \\
        & - \sum_{\ell=0}^\infty \frac{1}{\pi(2\ell+1)^2}\sin\left(\frac{B}{2}(2\ell+1)\pi\right) \cosh(2(2\ell+1)\theta)\;.
    \end{split}    
\end{equation}
Comparing to \eqref{15} we identify the coefficients (we take the mass scale $m=1$)
\begin{equation}
    \begin{split}
        \beta_0 = & - \frac{1}{2}\log(2)\;, \\
        \beta_{2\ell+1} = & - \frac{\cos\left(\frac{B}{4}(2\ell+1)\pi\right) + (-1)^\ell \sin\left(\frac{B}{4}(2\ell+1)\pi\right)}{2\ell+1}\;, \\
        \beta_{4\ell} = & - \frac{\cos(B \ell \pi)}{4\ell}\;, \\
        \beta_{4\ell + 2} = & - \frac{\cos\left(\frac{B}{2}(2\ell+1)\pi\right) + B \cos\left(\frac{B}{4}(2\ell+1)\pi\right) }{2(2\ell+1)} - \frac{1}{\pi(2\ell+1)^2}\sin\left(\frac{B}{2}(2\ell+1)\pi\right)\;.
    \end{split}
\end{equation}
\begin{figure}
    \centering
\includegraphics[width=0.7\linewidth]{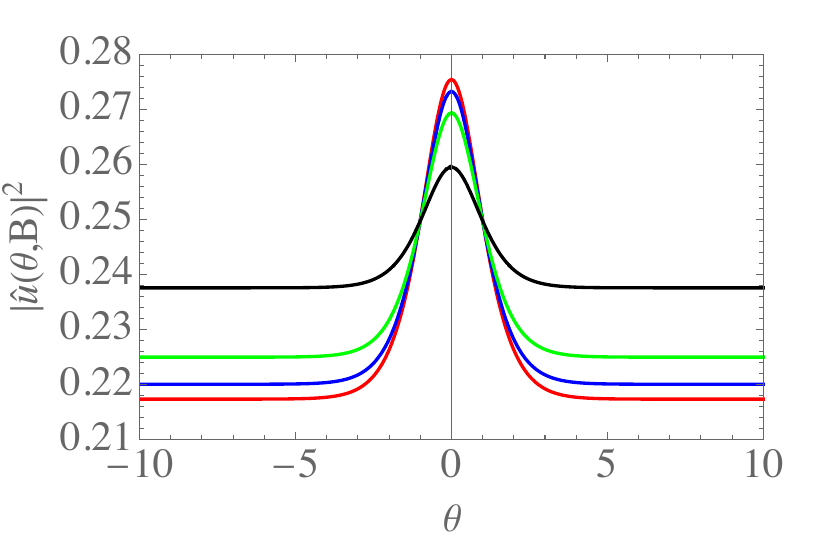}
    \caption{The absolute value squared of the function $\hat{u}(\theta,B):=\frac{u(\vartheta,B)}{i+\sinh\theta}$ with $u(\vartheta,B)=\exp{\omega(\vartheta,B)}$ evaluated numerically from (\ref{mainfor}). The colours correspond to different values of $B$: $B=1$ (red), $B=0.7$ (blue), $B=0.5$ (green) and $B=0.2$ (black).}
    \label{uhat}
\end{figure}
{ We note that the representation (\ref{mainfor}) is completely explicit and involves only elementary functions and a small number of special functions (dilogarithms). These are nonetheless functions that are efficiently implemented in all mathematical packages and therefore it is extremely easy and quick to evaluate (\ref{mainfor}) numerically with very high precision. We expect that this property will make our representation useful in the numerical evaluation of correlation functions and form factors.}

\section{More General Boundary Conditions}
\label{6}

We have just seen that the simplest boundary condition/reflection amplitude (\ref{thisR}) in the sinh-Gordon theory admits a new interpretation. In can be seen, at the level of the minimal form factor at least, as resulting from irrelevantly perturbing the boundary Ising model with fixed boundary conditions. The fixed boundary condition corresponds to taking the magnetic field $h$, which parametrizes all Ising boundary conditions, to infinity. Indeed, the most general reflection amplitude in the Ising model can be written as \cite{Ghoshal:1993tm}
\beq
R_x(\theta)= -[x]_\theta \left(-\frac{1}{2}\right)_\theta\,,
\label{RxIsing}
\eeq
where $x$ is related to the magnetic field $h$ as $\sin(\pi x)=1-\frac{h^2}{2m^2}$. 
There are two simple boundary conditions known as ``free" an ``fixed". They correspond to
\begin{itemize}
    \item  $h=0$ ($x=\frac{1}{2}$) with reflection amplitude
    \beq
    R_{\frac{1}{2}}(\theta)= \left(\frac{1}{2}\right)_\theta\,,
    \eeq
    This is the {\it{free boundary condition}}. As in the general case, this reflection amplitude has a pole at $\theta=\frac{i \pi}{2}$. This pole is dynamical, i.e. it changes position when changing the value of $h$.
    \item $h \to \infty$ ($x \to i \infty-\frac{\pi}{2}$), with reflection amplitude
    \beq
    R_{i \infty-\frac{\pi}{2}}(\theta)= \left(-\frac{1}{2}\right)_\theta
    \eeq
    this is called \textit{fixed boundary condition} and it is the simplest configuration. It corresponds to moving the pole away from the physical strip. Notice that in this case the factor $-[x]_\theta \to 1$ while in the free case it is non trivial. 
\end{itemize}
It is not difficult to generalise the construction of the minimal form factor (\ref{solutionShG}) to the case of generic reflection amplitudes (\ref{GenR}). One viewpoint is to consider the Ising field theory with generic boundary conditions itself as a ``perturbation" of the Ising field theory with fixed boundary conditions. Again, this is meant in the sense of how we compute the minimal form factor. This viewpoint allows us to both generalise the Ising and the sinh-Gordon results, to more general boundary conditions.  We discuss this below.

%Notice another important fact, while modifying a bulk amplitude with a CDD factor makes the seed theory interacting, modifying the boundary amplitude with a CDD factor does not: we are just connecting to the boundary some "flux" coming from the bulk. Exactly in the $\TTb$-deformation case this kind of connection was already made in the context of Ads3/CFT2 where the $\TTb$-deformed CFT in 2d is dual to a particular deformation of a 3-dimensional theory with a boundary. Already from this we can expect that the MFF in this case will not be exactly of the type \eqref{phi} but rather a "free version" of it. To show all of this, we just need to repeat the construction of the previous section to this case. 

\subsection{A $\TTb$ Picture of the Boundary Ising Model with Generic Boundaries}

%Another interesting viewpoint is that the minimal form factor of the Ising field theory, with generic boundary conditions can also be written in such a way as to represent a perturbation away from the Ising fixed boundary condition. 

Let $r_x(\theta)$ be the minimal form factor of the Ising field theory with generic boundary conditions, corresponding to the reflection amplitude \eqref{RxIsing}. The minimal form factor should be a modified version of the solution for fixed boundary conditions such that
\beq
r_x(\theta)= r_{\rm fixed}(\theta)\varphi_x(\theta),
\eeq
which implies 
\beq
\varphi_x(\theta)=-[x]_\theta \varphi_x(-\theta)\,, \qquad -[x]_\theta\varphi_x(\theta)=\varphi_x(2 \pi i -\theta)\,.
\eeq
The task is now to compute the new function $\varphi_x(\theta)$. 
Employing the standard integral representations that can be found in many places, such as \cite{BFF}, we have that 
\beq
-[x]_\theta= (x)_{\theta}(1-x)_{\theta}= \exp{\left\{ 2 \intop_0 ^{\infty}\frac
{dt}{t} \frac{\sinh{\frac{t \theta}{i \pi}}}{\sinh^2{t}} \left( \sinh{t x}+\sinh{t(1-x)} \right)\right\}}
\eeq
so we can write
\beq
\varphi_x(\vartheta)=\exp{ \left\{ 2 \intop_0 ^{\infty}\frac
{dt}{t} \frac{ \left( \sinh{t x}+\sinh{t(1-x)} \right)\cosh\frac{t}{2}}{\sinh^2{t}} \left( 
1- \cos{\frac{t \vartheta}{\pi}}
\right)\right\}}
\eeq
where we again use the variable
 $\vartheta= \frac{i \pi}{2} - \theta$. As before we take the logarithmic derivative
\beqa
\frac{d}{d \vartheta}\log \varphi_x(\vartheta) &=& \frac{2}{\pi} \intop_0 ^{\infty}\frac
{dt}{t} \frac{ \left( \sinh{t x}+\sinh{t(1-x)} \right) \cosh{\frac{t}{2}}}{\sinh^2{t}} \sin{\frac{t \vartheta}{i \pi}} \nonumber \\
&=& h(\vartheta,4x)+h(\vartheta,4-4x)\,,
\label{sum4x}
\eeqa
where $h(\vartheta,x)$ is the same function defined earlier in (\ref{hdef}). We can therefore use the same formula (\ref{hint}) to write
\beqa
h(\vartheta,4x)&=& \frac{\vartheta \sin{2 \pi x} - \pi \sin{\pi x}\sinh{\vartheta} + \pi x \sinh(2\vartheta)}{\pi (\cosh(2\vartheta)+\cos{2\pi x })}.
\eeqa 
Until now everything is pretty much the same as in previous sections. However, the sum (\ref{sum4x}) simplifies greatly, so that after integration, we have simply 
\beq
\log \varphi_x(\vartheta)= \intop d \vartheta [h(\vartheta,4x)+h(\vartheta,4-4x) ]= \log{\left( \cosh{\vartheta}+\sin{\pi x} \right)}+c\,.
\eeq
The constant is easily fixed to $c=-\log{\left( 1+\sin{\pi x} \right)}$ so as to ensure that $\log \varphi_x(0)=0$. Therefore, writing everything back in terms of $\theta$ we get
\beq
\varphi_x(\vartheta)= \frac{\sin{\pi x} - i \sinh{\theta}}{1+\sin{\pi x}}= \frac{\sin{\pi x} - i \sinh{\theta}}{\sin{\pi x} + i \sinh{\theta}}\, \frac{\sin{\pi x} + i \sinh{\theta}}{1+\sin{\pi x}}\,,
\eeq
or 
\beq
\log \varphi_x(\vartheta)= \log(-[x]_\theta)+ \log{(\sin{\pi x} + i \sinh{\theta})}-\log{(1+\sin{\pi x})}\,.
\eeq
This gives the generic factor that has to be added any time the boundary condition is changed. In particular, for $x=\frac{1}{2}$ we can obtain the modification of the minimal form factor with fixed boundary conditions that corresponds to free boundary conditions in the Ising case. Since the factor $\varphi_x(\vartheta)$ accounts for the contribution to the minimal form factor of a generic square block $-[x]_\theta$ in the reflection amplitude, it can also be adapted to deal with the blocks $[\frac{F-1}{2}]_\theta$ and $[\frac{E-1}{2}]_\theta$ in the sinh-Gordon amplitude (\ref{GenR}). 

\section{Conclusion and Outlook}
\label{3}
In this paper we have studied boundary IQFTs perturbed by $\TTb$ and higher spin irrelevant operators. We discussed how the deformation of the two-body scattering matrix (\ref{s}) propagates to a deformation of the reflection amplitudes off the boundary (\ref{5})-(\ref{8}) and how these give rise to a deformation of the one-particle minimal form factor (\ref{even})-(\ref{15}). While the deformation of the reflection amplitudes had already been discussed in \cite{Jiang:2021jbg,Brizio:2024doe}, this work initiates the study of form factors of irrelevantly perturbed boundary theories. 

We find that the form factor deformation is very similar to the bulk case. A further analogy is that this deformation suggests a factorised minimal form factor structure that is also reproduced for more standard boundary IQFTs. We show this to be the case for the sinh-Gordon model with Dirichlet boundary conditions, which we discuss in detail here. Through a computation which is analogous to that presented in \cite{MFF} we show that the boundary one-particle minimal form factor admits a new representation which forgoes integrals or infinite products, is very explicit and numerically efficient. In this representation, the boundary sinh-Gordon model with Dirichlet boundary condition can be interpreted as the Ising field theory with fixed boundary conditions in the presence of infinitely many irrelevant perturbations with specific coupling constants which are functions of the sinh-Gordon coupling $B$. The effect of adding irrelevant boundary operators in the Ising and sine-Gordon models was studied in  \cite{saleur}. Indeed, similar to our case and to the results of \cite{Jiang:2021jbg,Brizio:2024doe}, it was shown that such perturbations induce deformations of the reflection amplitudes which modify the UV properties of the theory. A similar conclusion has also been reached in the context of gravity, where it has been shown that $\TTb$ deformation of 2D conformal field theory can be seen as coming from a modification of the boundary conditions in a 3D (Chern-Simons) gravity theory \cite{Llabres:2019jtx,Ebert:2022ehb}.

Our construction easily generalises to other models and boundary conditions. More importantly, it should now be possible to progress to constructing non-minimal form factor solutions, as done for the bulk case in \cite{PRL,longpaper}. We hope to return to this problem in the near future.
\medskip

\noindent {\bf Acknowledgement:} Fabio Sailis is funded through a PhD studentship from the School of Science and Technology of City St George's, University of London, which he gratefully acknowledges.

\justifying
\bibliography{Ref}

\end{document}